\documentclass[fleqn,usenatbib]{mnras}
\usepackage{hyperref}
\usepackage{newtxtext,newtxmath}

\usepackage[T1]{fontenc}
\usepackage{ae,aecompl}

\usepackage{graphicx}	
\usepackage{amsmath}	
\usepackage{amssymb}	
\usepackage{xcolor} 	

\usepackage[figure,figure*]{hypcap}

\newcommand{\edit}[1]{\textbf{\textcolor{purple}{#1}}}

\newcommand\Msun{M_\odot}
\newcommand\Mjup{M_\mathrm{J}}

\newcommand\Lx{L_\mathrm{x}}

\newcommand\Myr{\mathrm{Myr}}
\newcommand\Gyr{\mathrm{Gyr}}

\newcommand\ergs{\mathrm{erg\,s^{-1}}}
\newcommand\ergscm{\mathrm{erg\,s^{-1}\,cm^{-2}}}
\newcommand\au{\mathrm{au}}
\newcommand\pc{\mathrm{pc}}

\newcommand\Chandra{\textit{Chandra}}
\newcommand\Rosat{ROSAT}
\newcommand\XMM{\textit{XMM-Newton}}
\newcommand\twomass{\textsc{2mass}}
\newcommand\simbad{\textsc{Simbad}}

\newcommand\Gaia{\textit{Gaia}}

\title[X-ray photoevaporation \& the orbital distribution of giant planets]{The imprint of X-ray photoevaporation of planet-forming discs on the orbital distribution of giant planets}

\author[Monsch et al.]{Kristina Monsch$^{1}$, Barbara Ercolano$^{1,2}$, Giovanni Picogna$^{1}$, Thomas Preibisch$^{1}$, \and Markus Michael Rau$^{3}$
\\ 
$^1$Universit\"ats-Sternwarte, Ludwig-Maximilians-Universit\"at M\"unchen, Scheinerstr. 1, 81679 M\"unchen, Germany\\
$^2$Excellence Cluster Origin and Structure of the Universe,
Boltzmannstr.2, 85748 Garching, Germany\\
$^3$McWilliams Center for Cosmology, Department of Physics, Carnegie Mellon
University, Pittsburgh, PA 15213, USA\\
{\tt E-mail: kristina.monsch@gmail.com}
} 

\date{Accepted 2018 December 4. Received 2018 November 30; in original form 2017 December 6}

\pubyear{2018}

\begin{document}
\label{firstpage}
\pagerange{\pageref{firstpage}--\pageref{lastpage}}
\maketitle

\begin{abstract}

High energy radiation from a planet host star can have strong influence on the final habitability of a system through several mechanisms. In this context we have constructed a catalogue containing the X-ray luminosities, as well as basic stellar and planetary properties of all known stars hosting giant planets ($ > 0.1~\Mjup$) that have been observed by the \Chandra\ X-ray Observatory, \XMM\ and/or \Rosat. Specifically in this paper we present a first application of this catalogue to search for a possible imprint of X-ray photoevaporation of planet-forming discs on the present-day orbital distribution of the observed giant planets. We found a suggestive void in the semi-major axis, $a$, versus X-ray luminosity, $\Lx$, plane, roughly located between $a\sim0.05$--$1~\au$ and $\Lx\sim10^{27}$--$10^{29}~\ergs$, which would be expected if photoevaporation played a dominant role in the migration history of these systems. However, due to the small observational sample size, the statistical significance of this feature cannot be proven at this point.
\end{abstract}



\section{Introduction}
\label{sec:intro}

Recent surveys have shown a vast diversity of extra-solar planets, raising the question of how some systems end up looking like our own and/or become hospitable for life. Hints to explaining the diversity of planetary systems may be in the understanding of the statistical trends that are now emerging from the recent wealth of observational data. 

One of these trends is the semi-major axis distribution of gas giants, which shows mountains and deserts, i.e. regions of over- and underpopulation. Understanding what controls the architecture of the new-born systems is crucial to assessing their habitability. As well as the location of terrestrial planets (i.e. inside or outside the habitable zone of their host star), the location of giant planets in a system is also of high importance to habitability, due to their central role in the delivery of volatiles to terrestrial planets \citep[e.g.,][]{QuintanaLissauer2014, Sanchez+2018} or stopping the influx of pebbles from the outer disc, possibly preventing the early formation of terrestrial planets \citep[e.g.,][]{Ormel2017}.

The majority of exoplanets are most likely not formed in-situ. A clear evidence of this is the population of so-called ``hot Jupiters'', gas giants with very short orbital periods, corresponding to semi-major axes $\lesssim0.1~\au$. 
The finding of a hot Jupiter around V830, a young T~Tauri star \citep{Donati+2016}, suggests that giant planets can migrate inwards in the planet-forming disc in less than two million years. As the detection rate of hot Jupiters around young stars ($6\,\%$) seems to be higher than the detection rate around mature stars ($1\,\%$) this suggests that planet-disc interactions are the main driver of planetary migration \citep{Donati+2016, Yu+2017}. 
However, the orbital characteristics of a system are, in fact, influenced by different processes that operate at different times. At the early stages of planet evolution, young planets are subject to torques exerted on them by the gas in the protoplanetary disc, which typically results in their inward migration \citep[e.g.,][]{GoldreichTremaine1980, LinPapaloizou1986, KleyNelson2012}. 
After a few million years, once the gas in the disc is finally dispersed, disc-driven migration stops. 
The timescales for migration of giant planets formed at radii of a few astronomical units are comparable to the gas disc lifetimes of a few million years \citep{HaischLadaLada2001}, suggesting that without a mechanism to halt this process, many planets would be pushed to the very inner regions of the disc and perhaps even be lost onto their host star. Indeed, planetary population synthesis models always include simplified prescriptions for eventually dissipating the disc \citep[e.g.,][]{Benz+2014, Mordasini+2015, Mordasini2018}. 
Without a parking mechanism one would expect to observe a much larger fraction of hot Jupiters. Recent exoplanet surveys, however, do not show an over-abundance of hot Jupiters. On the contrary, giant planets seem to clump preferentially in a region between $\sim1$--$2~\au$. 
What determines this peak in the distribution of giant planets is a matter of strong interest in the field today. 
\citet[][henceforth AP12]{AlexanderPascucci2012} suggested that internal photoevaporation driven by the host star could provide a natural parking radius for migrating giant planets, which was confirmed by numerical simulations of \citet[][henceforth ER15]{ER15} and \citet{Jennings+2018}.
These authors additionally showed that different photoevaporation profiles, i.e. the radially-dependent mass loss rate due to different photoevaporation models, have a dramatic influence on the final giant planet distribution for a given exoplanet population. 
While for most systems the orbits at the end of the gas disc lifetimes will still be dynamically evolving, this initial migration phase is extremely important in determining the dominant exoplanet distribution characteristics, especially if it leads to non-tightly packed systems after disc dispersal.

While the influence of the photoevaporation process on the initial stages of planetesimal formation is somewhat debated in the literature \citep[e.g.,][]{ErcolanoJennings+2017b}, its effect at later stages, when planetary cores have already been formed and are migrating through the disc has been demonstrated. If, as suggested by ER15, disc photoevaporation driven by stellar X-rays \citep{Ercolano+2009, Owen+2010, Owen+2011, Owen+2012} is affecting the observed distribution of giants, one may expect to observe a signature of this effect in the giant planet semi-major axis versus host star X-ray luminosity plane.
In this work we have gathered information on the X-ray properties of giant planet-hosting stars from the three most important X-ray telescopes: the \textit{Chandra} X-ray Observatory, \textit{XMM-Newton} and ROSAT. These were combined with basic stellar and planetary properties from the \textit{Extrasolar Planets Encyclopaedia}\footnote{\url{http://www.exoplanet.eu}} \citep{exoplanet.eu} to provide an extensive catalogue that can be used for investigating various processes between giant planets and their host stars.

Several previous studies already investigated different correlations between stellar and planetary properties, especially in context of the possible enhancement of stellar magnetic activity due to the presence of hot Jupiters \citep[e.g.][]{Lanza2008}. Early studies by \citet{Kashyap+2008} and \citet{Scharf2010} found that hot Jupiter hosts systematically show enhanced X-ray activity. However, \citet{Poppenhaeger+2010} constructed a sample of 72 planet-hosting stars within $30~\pc$ distance from the Sun by using \Rosat\ and \XMM\ data with which \citet{PoppenhaegerSchmitt2011} were able to show that there is no detectable influence of hot Jupiters enhancing the X-ray activity of their host stars. Subsequently, the study of \citet{Miller+2015} and \citet{Hinkel+2017} improved on several aspects, especially in terms of sample size. By including \Chandra\ observations, they were able to extend their sample to fainter and more distant stars, increasing the sensitivity and reliability of their correlation tests. 
The analysis performed in our work is, however, significantly different to these previous studies that mainly focused on hot Jupiter systems. 
We study the possible connection between the architecture of giant planet systems and the X-ray emission of their parent stars, using observations of their present-day luminosities, and compare our results to predictions from numerical models of the early ($<10~\Myr$) evolution of these systems.
Our increased sample size, largely constituted of deep \Chandra\ observations, enables us to better assess the possible influence of X-ray driven photoevaporation onto the final configuration of giant planets, with special attention to the orbital properties of ``warm Jupiters'', i.e. gas giants with orbital periods between 10 to 200 days.

As we use the present-day ($\sim\Gyr$) X-ray luminosities of planet hosts to relate to the evolution of the protoplanetary disc driven by the X-ray properties of the same stars at much younger ages ($\sim\Myr$), it is important to consider what is currently known about the origin and evolution of the stellar X-ray emission around low-mass stars. This is summarised in \S\,\ref{sec:x-ray}. The data retrieval and the resulting catalogue, which is included in the online supplementary material, is described in \S\,\ref{sec:observations}. The results are discussed in \S\,\ref{sec:results} and a brief summary is given in \S\,\ref{sec:summary}.


\section{Origin and Evolution of the X-Ray Emission}
\label{sec:x-ray}

In this work we study the connection between the orbital configuration of extrasolar giant planets and the X-ray emission of their parent stars. 
The connection between these two aspects may stem from the evolutionary processes during the first few million years of the system, when the planets are still embedded in the protoplanetary disc, whose evolution and final dispersal are driven by the stellar irradiation at those young ages. 
Observationally, however, we can only measure the present day X-ray luminosities of the planets' host stars. 
It is thus important to discuss not only the origin of the X-ray emission but also how it evolves over time.

\subsection{Origin of stellar X-ray emission}
\label{sec:xray_origin}

The X-ray emission of late-type stars with spectral types ranging from F to M originates from a hot, magnetically confined plasma in the stellar corona \citep[e.g.,][]{Vaiana+1981, Guedel2004, Jardine+2006, Reale2014}, resulting from magnetic dynamo activity produced in a boundary layer between the radiative core and the convective envelope, known as \textit{tachocline} \citep[e.g.,][]{Wright+2011}.
Although there are still some open questions about the exact nature of the magnetic dynamo at work \citep[e.g.,][]{WrightDrake2016} and especially about the origin of activity saturation for very rapidly rotating stars \citep[e.g.,][]{Kitchatinov2015}, it is well established that magnetic dynamo activity is primarily determined by the stellar rotation rate.

Numerous X-ray observations obtained during the last decades have clearly established that young stars in all evolutionary stages from protostellar to zero age main sequence (ZAMS) show highly elevated levels of X-ray activity \citep{Preibisch+1996, FeigelsonMontmerle1999, PreibischZinnecker2002, FavataMicela+2003, Preibisch+2005, Preibisch+2014}. 
Typical X-ray luminosities of solar mass young stellar objects (YSOs) during the first $\sim100~\Myr$ are about $10^{30}$--$10^{31}~\ergs$, i.e. up to $\sim 10^4$ higher than of the current Sun. The temperatures of the X-ray emitting plasma of young stars are typically 10 to 20~MK, i.e. about ten times higher than in the solar corona.

Two fundamental results derived from particularly deep X-ray observations of young clusters are that \textit{(i)} X-ray luminosity scales with stellar mass as $\Lx \propto M^{1.44}$ to $M^{1.54}$ \citep{Preibisch+2005, Guedel+2007_Taurus} and \textit{(ii)} the X-ray luminosity of T-Tauri stars (TTS) is approximately constant for ages up to $10~\Myr$, and then decreases following a power-law dependence of $\Lx \propto t^{-0.75}$ \citep{PreibischFeigelson2005}. In light of the above described dynamo theories, the high X-ray activity levels of young stars are thought to be an ultimate consequence of their fast rotation during the first few $100~\Myr$ of their lives \citep[e.g.,][]{AlexanderPreibisch2012}. The rotational evolution of stars, discussed in the next paragraph, explains the time evolution of their X-ray activity.

\subsection{Evolution of stellar rotation}
\label{sec:xray_evolution}

During their first few million years, YSOs are generally surrounded by circumstellar accretion discs, in which the accretion of matter from the disc onto the star is magnetically controlled \citep[cf.][for reviews]{Bouvier+2007, Hartmann+2016}. In this scenario, circumstellar discs are connected to their host star via magnetic field lines that lead to their truncation at or near the co-rotation radius, which lies typically, depending on the rotation period of the star, at a few ($\sim$3--5) stellar radii \citep{Bouvier+2007, Preibisch2012}.
This magnetic coupling regulates the rotation rate of the star, which in turn affects the stellar X-ray emission. 
In particular, those field lines, which connect the stellar surface to parts of the disc that lie beyond the co-rotation radius, will tend to slow down the stellar rotation rate, as here the rotation frequency of the disc is lower than the rotation frequency of the star \citep[cf.][and references therein]{Bouvier+2007}.
This kind of magnetic coupling is assumed to be the explanation of the rather moderate rotation rates observed from young stars with ages between $\sim 1~\Myr$ and a few Myr. 

When the disc has been dissipated after a few Myr, the braking effect of the magnetic coupling between the star and the disc declines and finally comes to a halt, and thus the rotation of the young star speeds up as a consequence of its continuing contraction. 
The highest rotation rates are established around the age of $\sim 30$--$50~\Myr$ for solar-type stars, i.e. around the time the star reaches the ZAMS and its contraction stops. 
At later times, the rotation rates of young stars decrease steadily, mainly caused by the loss of angular momentum from stellar winds \citep[e.g.,][]{Guedel2007, GalletBouvier2013, GalletBouvier2015, Tu+2015}. 
The rotational evolution therefore depends to a large degree on the time scale over which the star is magnetospherically coupled to, and braked by the disc \citep[e.g.,][]{Montmerle+2000}. 
Consequently, higher rotation frequencies are reached for such stars that disperse their accretion disc more quickly than for stars for which the braking effect of the magnetospheric coupling lasts longer.
This therefore explains the wide range of observed rotation velocities for young stars with ages between $\sim 30$--$100~\Myr$ and the resulting scatter in X-ray luminosity for stars of these young ages. While the decline in mean X-ray luminosity is small for the first $100~\Myr$ compared to the observed spread in $\Lx$ for a given age \citep[cf.,][]{PreibischFeigelson2005}, at later ages, when the stellar rotation rates decrease, the range in rotational velocities (and consequently $\Lx$) gets smaller before converging to the very low values typical for older main-sequence stars with ages of a few Gyr, such as our Sun. This behaviour closely follows the well-known Skumanich's relationship $\Omega_* \propto t^{-0.5}$ \citep{Skumanich1972}.

However, up to ages of about 1--$2~\Gyr$, stars with the highest rotation rates during their early evolution are still among the fastest rotators in their current age group, i.e. the rank in rotation rates is conserved during that timescale \citep{GalletBouvier2013, GalletBouvier2015, Johnstone+2015, Tu+2015}.
This therefore justifies using the present-day observed X-ray luminosities of planet-hosting stars younger than about 1--$2~\Gyr$ as a proxy for the X-ray luminosity experienced by the same planet-disc systems at the time of giant planet migration and disc dispersal.

\section{Observational Analysis}
\label{sec:observations}

We used the \textit{Extrasolar Planets Encyclopaedia}\footnote{\url{http://www.exoplanet.eu}} \citep{exoplanet.eu} to get a collection of all confirmed and unconfirmed extrasolar planets as of September 2017. In this project, we focus solely on the early dynamical evolution of giant planets that undergo so-called type~II migration, therefore a lower threshold of $0.1~\Mjup$ was set for the planetary mass. In addition, all planets detected by gravitational microlensing or pulsar timing were excluded from this analysis.
Since the target coordinates listed in the encyclopaedia are often unreliable, we used the stellar coordinates as listed in the \simbad\ Astronomical Database \citep{simbad} for the cross-matching.
Further, we used the catalogue generated by \citet{Bailer-Jones+2018}\footnote{\url{http://www.mpia.de/~calj/gdr2_distances/main.html}}, in which accurate distance estimates inferred from the most recent \Gaia\ Data Release 2 are given \citep[Gaia DR2,][]{GAIA_mission, Gaia_DR2}.

\subsection{\Chandra\ X-ray Observatory}
\label{sec:chandra}

We matched the stellar positions with the footprints of all ACIS and HRC observations listed in the \Chandra\ archive\footnote{\url{http://cda.harvard.edu/chaser/}}. To cover the entire field-of-view of both the ACIS and the HRC instruments, a cone search radius of $12'$ and $22'$ was used, respectively. 
We checked the accuracy of the astrometry in the \Chandra\ data by comparison of X-ray source positions with the \twomass\ Point Source catalogue \citep{2MASS}.
At the celestial position of the planet host star in the \Chandra\ image, we ran the \texttt{CIAO} script \texttt{srcflux}\footnote{\url{http://cxc.harvard.edu/ciao/ahelp/srcflux.html}} \citep{ciao} in order to determine the broad band ($0.5$--$7.0~\mathrm{keV}$) X-ray fluxes for X-ray detected stars and corresponding upper limits for undetected ones. For this purpose we used the thermal Bremsstrahlung model\footnote{The XSPEC model function used in this work was \texttt{xsbremss}. A detailed description of all available model functions in \texttt{CIAO} can be found at \url{http://cxc.harvard.edu/sherpa/ahelp/xs.html} and \url{https://heasarc.gsfc.nasa.gov/docs/xanadu/xspec/manual/node143.html}.} implemented in \texttt{CIAO} with a plasma temperature of $kT=0.5~\mathrm{keV}$ and a galactic hydrogen column density of $N(\mathrm{H})=10^{20}~\mathrm{cm^{-2}}$.

\subsection{\XMM}
\label{sec:xmm}

Count rates for \XMM\ source detections were retrieved from the seventh data release of the Third \XMM\ Serendipitous Source Catalogue \citep[3XMM-DR7,][]{Rosen+2016_3XMM-DR8}.
For undetected sources, upper limits on the source count rates were determined using the web tool FLIX\footnote{\url{http://www.ledas.ac.uk/flix/flix_dr5.html}}. This tool applies the algorithm described by \citet{Carrera+2007} to estimate the upper limits of the band~8 ($0.2$--$12.0~\mathrm{keV}$) source fluxes using the energy conversion factors (ECFs) listed in \citet{Mateos+2009}.
To allow for the direct comparison between the source fluxes of different X-ray telescopes, the band~8 \XMM\ fluxes were converted to $0.5$--$7.0~\mathrm{keV}$ to match the \Chandra\ ACIS broad band fluxes using the web tool WebPIMMS\footnote{\url{https://heasarc.gsfc.nasa.gov/cgi-bin/Tools/w3pimms/w3pimms.pl}} with $kT=0.5~\mathrm{keV}$, $N(\mathrm{H})=10^{20}~\mathrm{cm^{-2}}$ applying the same spectral model as described in \S\,\ref{sec:chandra}.

\subsection{ROSAT}
\label{sec:rosat}

In order to further increase the sample of planet-hosting stars observed in X-rays, several catalogues based on the \Rosat\ All-Sky Survey \citep[RASS, ][]{rosat_brightsources, rosat_faintsources, rosat_2nd_pspc, rosat_1RXH} including the most recent second data release from 2016 \citep{rosat_2RXS} have been searched. These catalogues provide the count rates for detected stars, which we then converted into source fluxes using the same spectral model as discussed before. These were also converted to $0.5$--$7.0~\mathrm{keV}$, as described in \S\,\ref{sec:xmm}, to match the \Chandra\ broad band fluxes.
Following \citet{Miller+2015}, we estimated approximate RASS flux upper limits using the typical detection limit of the RASS Faint Source Catalogue of $10^{-13}~\ergscm$. Due to the shallowness of this all-sky survey, this resulted in very crude estimates of the flux upper limits, which are several orders of magnitude larger than those estimated from \Chandra\ observations. Therefore, we do not include these upper limits in our catalogue.

\subsection{Final catalogue}
\label{sec:catalog}

Some of the targets were observed by multiple X-ray telescopes. In such cases we prioritised \Chandra\ over \XMM\ over \Rosat\ observations in our analysis, 
as \Chandra\ offers the highest angular resolution and thus very good point source sensitivity. For stars that have been observed multiple times by a given telescope, we calculated the average flux and used this value in the analysis. For undetected X-ray sources, we used the lowest upper limit, as this corresponds to the tightest constraint on the actual source flux. 

Stellar X-ray luminosities and their corresponding errors were then determined using the distance measurements either given in the \textit{Extrasolar Planets Encyclopaedia} or from \textit{Gaia} DR2 \citep{Bailer-Jones+2018, Gaia_DR2}, if available. 
While errors for \Chandra\ data points correspond to the $\pm1\,\sigma$ quantiles of the source fluxes, corresponding errors for \XMM\ and \Rosat\ were calculated using the stated errors in the respective catalogue. Other uncertainties, such as in the distance or the fixed energy conversion factors were not taken into account. Therefore we expect the true errors to be larger, in the order of 20\,\%--30\,\% \citep[cf.][]{Kashyap+2008, Poppenhaeger+2010, Miller+2015}


We further collected measurements of the stellar rotation velocity, $v\sin{i}$, for the host stars in our sample by cross-matching those with the catalogues compiled by \citet[][\href{http://vizier.u-strasbg.fr/viz-bin/VizieR?-source=J/ApJ/743/48}{VizieR ID J/ApJ/743/48}]{Wright+2011} and \citet[][\href{http://vizier.u-strasbg.fr/viz-bin/VizieR?-source=III/244}{III/244}]{Glebocki+2005} using a search radius of $3''$ and $5''$, respectively (due to high uncertainties in the source positions for the latter).

Finally, we omit the detections of unconfirmed, gap-opening planets\footnote{These are namely TW~Hya a/b, HD~100546~b, HD~98800~B b, KH~15D b and QS~Vir~b.} still embedded in a protoplanetary disc, as their existence is highly debated. For example, a recent study by \citet{Ercolano+2017c_TWHya} suggests a photoevaporative origin of the gap seen in the disc surrounding TW~Hya rather than a giant planet.

The current version of the catalogue therefore contains X-ray luminosities and planetary data for nearly 200 stars hosting giant planets, out of which 124 are X-ray detections and 70 upper limits.


\section{Results and Discussion}
\label{sec:results}

\begin{figure*}
\centering
\includegraphics[width=0.7\linewidth]{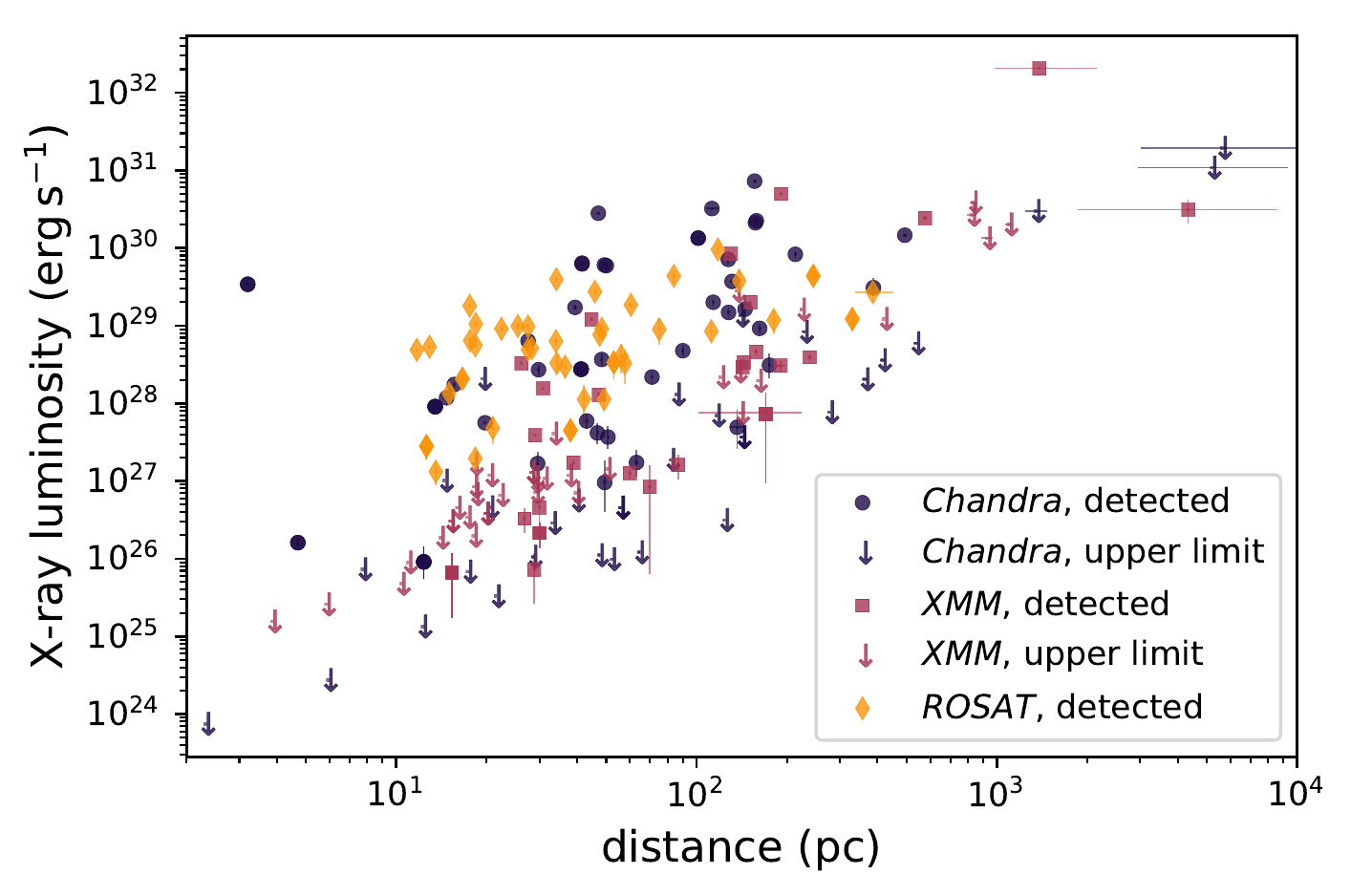}
\caption{X-ray luminosity as a function of distance for the planet-hosting stars in our catalogue. Upper limits are indicated as arrows and correspond to stars that have been observed but were not detected in a given X-ray observation. This plot excludes targets where only \textit{ROSAT} upper limits are available, as their X-ray luminosities have been crudely estimated using a fixed value for the stellar flux, as discussed in \S\,\ref{sec:rosat}.}
 \label{fig:Lx_vs_distance}
\end{figure*}

\subsection{Distribution of stellar and planetary properties in our catalogue}
\label{sec:properties}

Figure~\ref{fig:Lx_vs_distance} shows the X-ray luminosity distribution of all stars hosting planets with masses larger than $0.1~\Mjup$ in our catalogue as a function of distance. As expected, most detected systems are relatively nearby with distances $\lesssim 200~\mathrm{pc}$. Owing to \textit{Chandra}'s high sensitivity, it is however possible to infer meaningful upper limits on X-ray luminosities even for stars with distances up to $10^{4}~\mathrm{pc}$. While the most distant systems\footnote{SWEEPS-04 \& SWEEPS-11, detected within the \textit{Sagittarius Window Eclipsing Extrasolar Planet Search} near the Galactic Bulge \citep[SWEEPS,][]{Sahu+2006}.} in Figure~\ref{fig:Lx_vs_distance} are more likely to be an exception, X-ray information of systems up to several hundreds of parsecs can be routinely obtained. 
In particular the combination of shallow but abundant ROSAT and \textit{XMM-Newton} data with highly sensitive \textit{Chandra} observations enables us to obtain a rather complete sample and to cover a larger range of observed X-ray luminosities of giant planet hosts. 

\begin{figure*}
\centering
\includegraphics[width=0.9\linewidth]{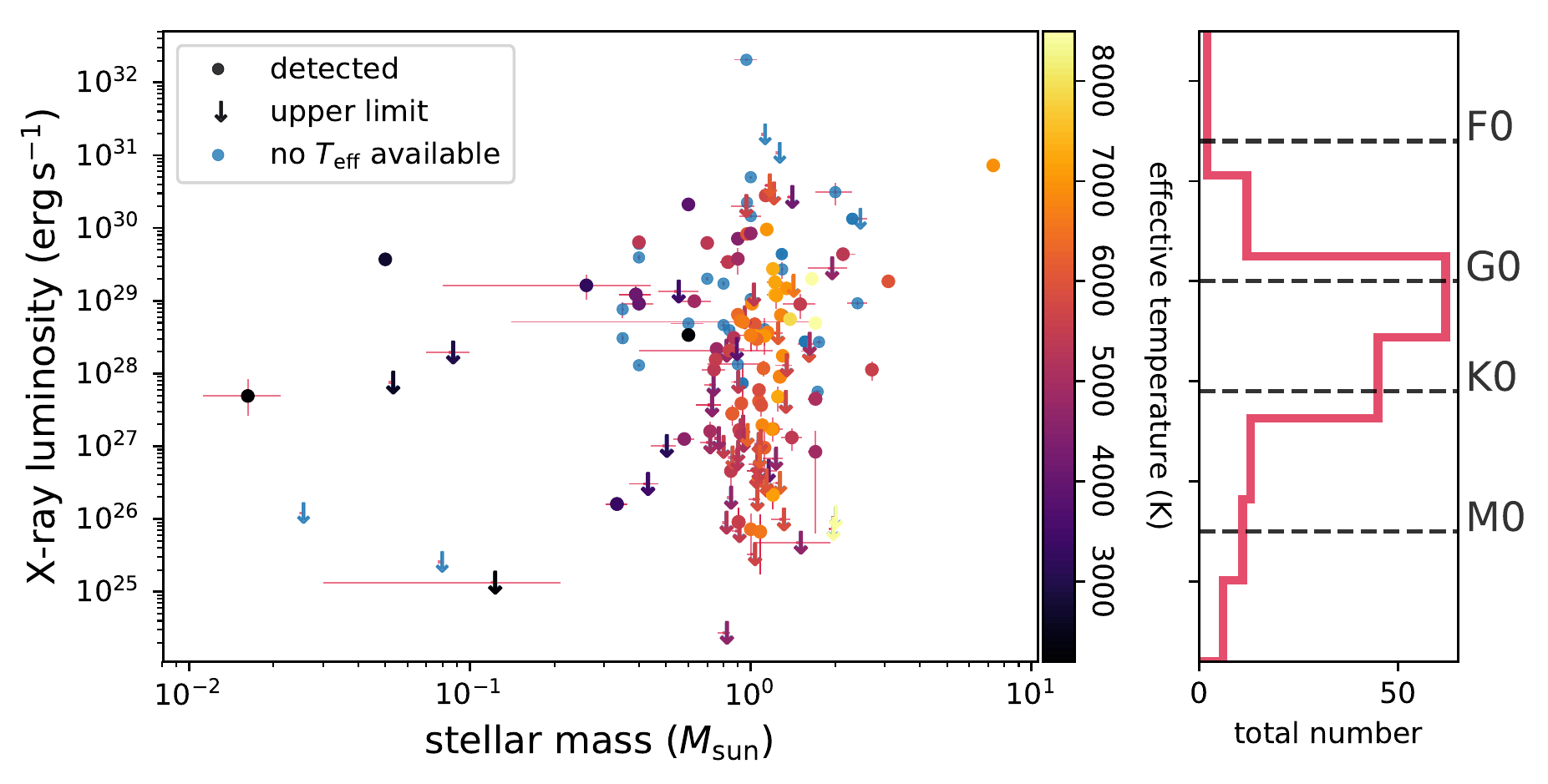}
\caption{X-ray luminosity of all host-stars in our catalogue as a function of their mass, with their effective temperature shown as colours. Upper limits are indicated as arrows and correspond to stars that have been observed but were not detected in a given observation. The histogram on the right shows the distribution of effective temperatures for these stars and their corresponding spectral types.}
 \label{fig:Lx_vs_Mstar}
\end{figure*}

Figure~\ref{fig:Lx_vs_Mstar} shows the resulting $\Lx$-vs-$M_{\mathrm{star}}$ distribution of the stars in our sample, with colours corresponding to the effective temperature of each star. 
One can immediately infer that this distribution peaks close to $\sim1~\Msun$, thus most stars being of solar-type with effective temperatures ranging from $4000$--$7000~\mathrm{K}$. These stars therefore mostly resemble G- and K-type stars, as can be seen in the histogram in the right panel of Figure~\ref{fig:Lx_vs_Mstar}.

\subsection{Observational biases and selection effects}
\label{sec:biases}

The main goal of this paper is to study whether X-ray driven photoevaporation has a detectable effect on the final semi-major axis distribution of giant planet systems and therefore we will only focus on the $\Lx$--vs--$a$ distribution of these systems in the following. 
However, as previously pointed out, our catalogue contains several other properties of these systems which could be used to study further correlations and effects, such as in the context of star-planet interactions or the influence of stellar irradiation on exoplanet atmospheres.

\begin{figure*}
\centering
\includegraphics[width=0.7\linewidth]{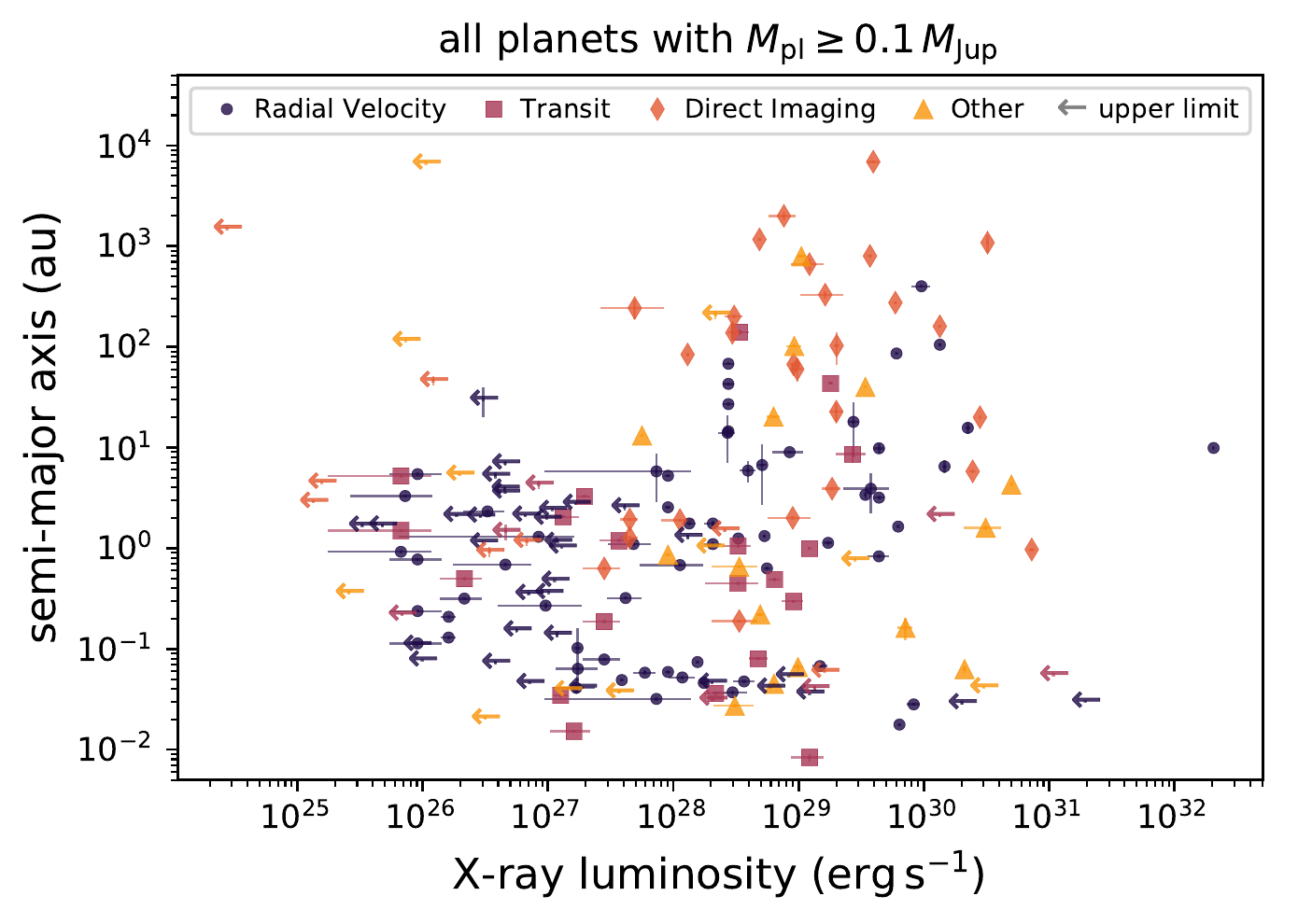}
\caption{Observed semi-major axis vs. X-ray luminosity distribution of stars hosting giant planets with masses above $0.1~\Mjup$, as discussed in \S\,\ref{sec:observations}. For multi-planet systems all planets with a mass of $0.1~\Mjup$ are plotted.
The colours highlight the different exoplanet detection methods for each system. Techniques that result in only very few detections (like astrometry or transit timing variation (TTV)) in our catalogue are summarised as 'Other'. Upper limits on X-ray luminosities are shown as arrows.}
 \label{fig:obs_detection_methods}
\end{figure*}

In Figure~\ref{fig:obs_detection_methods} we show the resulting X-ray luminosity distribution for all stars vs. the semi-major axes of their giant planets. 
The observed distribution covers several orders of magnitudes both in X-ray luminosities, $\Lx$, and in planet semi-major axes, $a$. Since the detection of extrasolar planets is subject to many observational biases, this results in several underpopulated regions, especially at large semi-major axes and low X-ray luminosities. 
A detailed description of many possible observational biases can be found in \citet{Kashyap+2008} and \citet{Miller+2015}, which will be further discussed in the following.

We have assigned different colours and symbols for each exoplanet detection method in Figure~\ref{fig:obs_detection_methods} to test whether they are introducing features in the overall distribution.
As expected, most planets ($\sim50\,\%$) have been detected by the radial velocity (RV) method, which is specifically sensitive to high-mass planets at small to intermediate orbits \citep[e.g.,][]{Wright2017_rv}.
While they are present over the entire X-ray luminosity range, they are mostly concentrated at lower values of $\Lx$, because X-ray bright stars are generally less frequently targeted by RV-surveys due to the high variability of the stellar emission. Such stars often correspond to young pre-main sequence stars whose high activity is impeding the identification of weak RV-signals (such as by low-mass or distant planets) within the stellar spectrum \citep[e.g.,][]{Jeffers+2014}.
However, new methods for detecting planets especially around young active stars are promising \citep[e.g.,][]{Jones+2017, Biddle+2018}, which could help to fill this underdense region in the $\Lx$--$a$ parameter space. 

Similar behaviour is found for planets detected by the transit method, which are, however, significantly less frequent than RV-detected planets in our sample. Due to the confined range of inclinations that is needed to observe planetary transits, this method is more restricted to close-in ($< 10~\au$) planets \citep[e.g.,][]{DeegAlonso2018_transit}. Further, most transiting planets are found around stars with intermediate to low X-ray activity ($\Lx \lesssim 10^{29}~\ergs$), as the presence of stellar spots on highly active stars does not only impede the detection of exoplanets but also the accurate determination of their physical properties \citep[e.g.,][]{Boisse+2011, Ballerini+2012}. 

For large semi-major axes (i.e. $\gtrsim10~\au$), the most sensitive planet detection method is direct imaging.\footnote{Also microlensing can detect planets with large semi-major axes, however, since follow-up observations are often impossible, their orbital properties are loosely defined \citep[cf.,][]{Gaudi_microlensing_2010, Winn_microlensing_2015}.}. This technique detects young planets more easily, since they are still releasing the accretion luminosity from their formation phase \citep[e.g.,][]{Spiegel2012} and are thus brighter in the infrared. Young planets are associated with younger, and therefore more X-ray bright stars (cf., \S\,\ref{sec:x-ray}), which explains why the upper right part of Figure~\ref{fig:obs_detection_methods} is more densely populated compared to the upper left region.

The remaining planets detected by astrometry or transit timing variation are summarised as `Other' in Figure~\ref{fig:obs_detection_methods}. Since they only make up a small fraction of our sample $(<10\,\%)$, we do not expect them to introduce any significant features in the $\Lx$--$a$-distribution. Nevertheless, this is expected to change with the recently published second data release of \textit{Gaia}, which will likely include many new exoplanet detections by astrometry and gravitational microlensing \citep{Perryman+2014_Gaia, KatzBrown2017_GaiaDR2}. 

Our catalogue includes in total 200 giant planets which have measured X-ray luminosities of their host stars. Compared to the \textit{Extrasolar Planets Encyclopaedia}, which included 1224 giant planets at the time the catalogue was prepared, this corresponds to 16\,\% of the total sample. If we restrict ourselves to wide-orbit planets only (i.e. $a > 5~\au$), our catalogue covers 44\,\% out of the planets from the exoplanet archive (i.e. 56 vs 126 planets). 
To check, if the stellar properties of the wide-orbit planet host stars differ significantly from those of the remaining sample, we have searched for trends and systematics in the distribution of different stellar properties but did not find any.
The upper left part of Figure~\ref{fig:obs_detection_methods} is therefore the only region where an observational bias is affecting all the detection methods. 
However, it is difficult to identify and account for all of these biases. Therefore, every analysis of possible correlations or trends within the data should be treated with caution, if the distorting effect of observational biases cannot be ruled out entirely \citep[cf.,][]{PoppenhaegerSchmitt2011}.

\subsection{Theoretical implications of X-ray photoevaporation onto the final locations of giant planets}
\label{sec:theory}

Before one can begin to search for possible features within the observed distribution of exoplanets, it is important to understand how X-ray photoevaporation may be affecting the dispersal of protoplanetary discs, which is eventually halting the planet migration process and therefore likely shaping their semi-major axis distribution in a characteristic manner. 
However, as discussed previously in \S\,\ref{sec:xray_evolution}, the rank in primordial X-ray luminosities is not expected to be conserved for stars with ages $> 2~\Gyr$. To be able to link the observed properties of the giant planet hosts to their X-ray luminosities at the time of disc dispersal, we will consider only `young' stars in the following, i.e. stars with ages below $2~\Gyr$.

\subsubsection{Predictions for expected features within the observations}
\label{sec:gap_shifts}

Mass loss by photoevaporation is concentrated at a given disc radial distance, around the so-called gravitational radius, which is roughly 1-2 au for solar type stars \citep{Hollenbach+1994, AA2009}. At this location a gap in the protoplanetary disc is formed when the mass loss rate due to photoevaporation exceeds the accretion rate through the disc. Thus naively one can expect to observe an underdensity of planets around the gravitational radius for a given range of X-ray luminosities for which a vigorous wind is expected ($\Lx$ of order 10$^{29}$--10$^{30}~\ergs$). As the disc is cleared from the inside out on a short time-scale, planets that are located outside of the photoevaporative gap are stopped, once the photoevaporative gap reaches their location. On the other hand, planets which are located inside the gravitational radius continue migrating, until the inner disc is dispersed due to viscous accretion. Assuming a broad range of initial conditions (such as the formation time and location of the planet or the strength of the photoevaporative wind), this results in an underdensity of giant planets located inside the gravitational radius \citep[cf.,][]{AlexanderPascucci2012}.

However, as already discussed in \S\,\ref{sec:xray_evolution}, the stellar X-ray luminosity decreases with age as a result of stellar spin-down. 
While protoplanetary disc dispersal occurs at much earlier stages of the stellar lifetime ($\lesssim10~\Myr$), the planet hosts in our catalogue mostly have ages of the order Gyr, when the X-ray luminosities are generally several orders of magnitude lower. 
It is thus to be expected that any features that are imprinted into the early $\Lx$--$a$-distribution at the time of disc dispersal would be shifted to lower X-ray luminosities with increasing age of the corresponding system. Note however that an exact mapping of present day X-ray luminosity to earlier times is very model-dependant and not trivial, and therefore this is not attempted in the present work.

Additionally, our catalogue contains stars with a range of stellar masses, which could smear out an underdensity feature, given that photoevaporation opens a gap at the gravitational radius $R_{\mathrm{g}}$, which is linearly dependant on the stellar mass. $R_{\mathrm{g}}$ is defined as
\begin{equation}
\label{eq:Rg}
R_{\mathrm{g}}=\frac{GM_{\star}}{c_{\mathrm{s}}^2}, 
\end{equation}
where $M_{\star}$ is the mass of the central star, $G$ the gravitational constant and $c_{\mathrm{s}}$ the sound speed of the gas \citep[][]{Hollenbach+1994}.
Thus a planet population with hosts of different stellar masses will produce deserts centered at several locations, depending on the gravitational radius of the members. Nevertheless, as solar-mass objects are the most abundant in our catalogue, one could still expect to observe an underdensity of objects at the location of the gravitational radius for solar-type stars (i.e. 1--$2~\au$).

\subsubsection{Dynamical evolution}
\label{ref:dynamical_evolution}

A fundamental assumption made in our study is that the observed configuration of planetary systems resembles the one established soon after the dispersal of the protoplanetary disc. This is somewhat justified by recent results \citep[cf.][]{Kipping2018}, that show that \textit{Kepler} multi-planetary systems present a highly significant deficit in entropy compared to a randomly generated population, suggesting that they indeed keep a memory of their initial state. Therefore, the early migration phase of newly formed planets in the disc is extremely important in setting the initial conditions for any later dynamical evolution after disc dispersal. 

\begin{figure}
\centering
\includegraphics[width=\columnwidth]{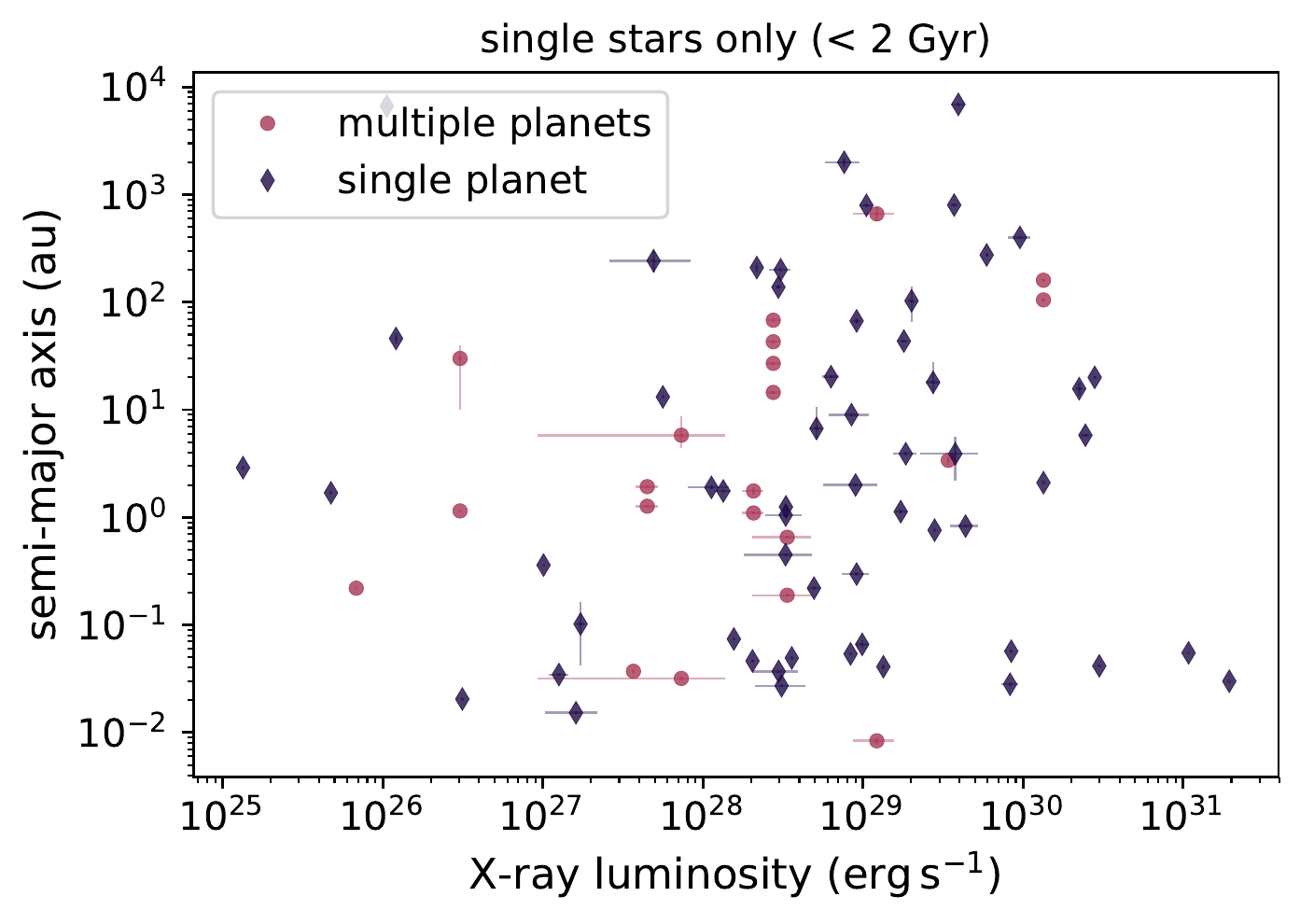}
\caption[]{Semi-major axis vs. X-ray luminosity distribution for young ($<2~\Gyr$), single stars hosting confirmed giant planets in single (circles) and multiple (diamonds) planetary systems. The latter also include planets that have a planetary companion with a mass lower than the minimum mass limit of $0.1~\Mjup$ in our catalogue.}
 \label{fig:multiplan}
\end{figure}

It is therefore tempting to interpret orbital parameters (e.g. eccentricity) for the gas giants in our sample in the context of dynamical evolution, both post- and pre-disc dispersal \citep{DawsonChiang2014, Huang+2016, PetrovichTremaine2016, AndersonLai2017, Sotiriadis+2017}. This is, however, a non-trivial task, first of all because eccentricities and inclinations are difficult to infer from transit detections \citep[e.g.,][]{Seager2003}, resulting in large uncertainties in their determination. Furthermore, the naive assumption that planetary eccentricities are damped by the gaseous disc, leading to circular orbits after disc dispersal is also an over-simplification. Indeed, a forming giant planet, which is able to carve a very deep gap in the disc, can be left on significantly eccentric orbits at the end of the disc dispersal phase \citep[e.g.,][]{Duffell2015}.

\citet{Wright2009} found that multi-planetary systems tend to have on average lower eccentricities \citep[see also][]{Huang+2016}.
This behaviour is expected since planets with lower eccentricities are dynamically more stable in the long-term.
Figure~\ref{fig:multiplan} shows the population of confirmed single and multiple planetary systems orbiting young (i.e. younger than $2~\Gyr$), single stars in our sample, where the latter also include planets that have a planetary companion with a mass below the minimum mass limit of $0.1~\Mjup$. No significant difference is shown in the distribution of single and multiple-planetary systems for $\Lx < 10^{29}~\ergs$, supporting our assumption that dynamical interactions between planets after disc dispersal do not change the overall picture. 
The lack of multiple-planetary systems for high $\Lx$ can be the result of faster disc photoevaporation, preventing the formation of multiple giant planet systems.

\subsection{Deserts in the $\Lx$-$a$-plane - signatures of X-ray driven photoevaporation?}
\label{sec:obs_vs_theory}

\begin{figure}
\includegraphics[width=\linewidth]{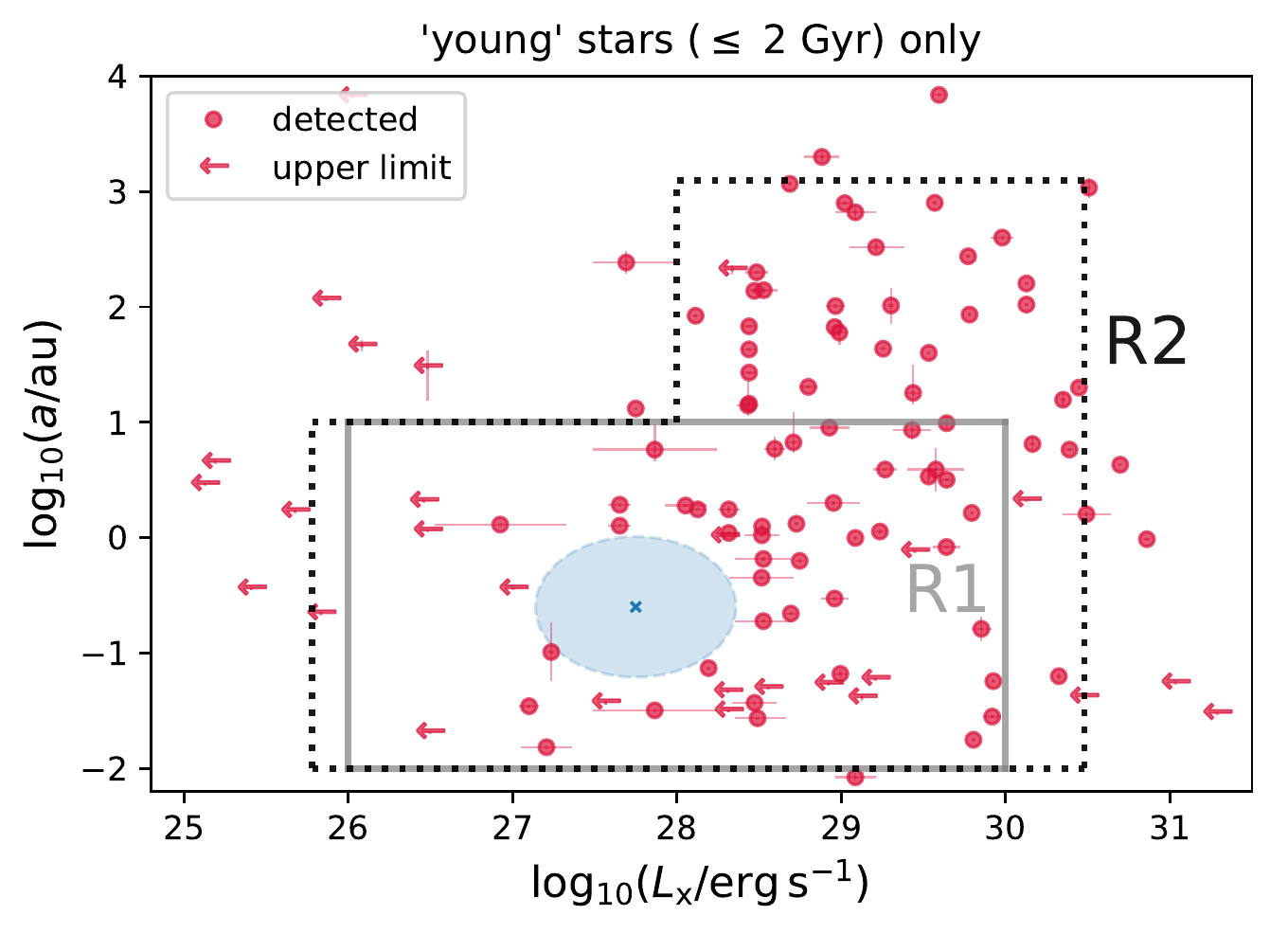}
\caption{Semi-major axis vs. X-ray luminosity distribution of detected (circles) and undetected (arrows) stars in our catalogue. The grey/dotted contours show the reference regions used in the statistical analysis to test the significance of the encircled voids as described in \S\,\ref{sec:obs_vs_theory}. The blue region highlights the void, in which no young stars are found.}
 \label{fig:results}
\end{figure}

Previous work by \citet{AlexanderPascucci2012} and \citet{ER15} suggests that signatures from X-ray driven photoevaporation in the form of under- and overpopulation are likely to be imprinted in the final semi-major axis distribution of giant planets. While dynamical evolution of the planets after disc dispersal can alter their final location in the system drastically, their initial conditions are however set by the early evolution and migration phase while the disc is still present. 
Looking at the observed distribution shown in Figure~\ref{fig:obs_detection_methods} ($\sim \Gyr$), we find several underdense regions that \textit{could} be related to X-ray driven photoevaporation creating distinct features in the distribution of planets. 
These are, for example, centred on $(\Lx, a)\sim(3\cdot10^{29}\,\ergs, 0.2\,\au)$, $(2\cdot10^{28}\,\ergs, 5\,\au)$ or $(10^{27}\,\ergs, 1\,\au)$.
Nonetheless, as pointed out in \S\,\ref{sec:biases}, our catalogue is subject to many observational biases and selection effects that can hardly be all accounted for. Especially the underdensities seen near the edges of the $\Lx$--$a$-distribution in Figure~\ref{fig:results} (which is similar to Figure~\ref{fig:obs_detection_methods}, however now only young stars (i.e. $<2~\Gyr$) are shown, as they are the only ones from which a memory of their primordial X-ray luminosity is to be expected, cf. \S\,\ref{sec:xray_evolution}) are most likely the result from observational biases rather than real physical effects, which is affirmed by the generally smaller number of data points in these regions. 
The only void in this distribution that cannot be readily explained by any observational biases or selection effects, is the desert centred on $[\log(\Lx/\ergs), \log (a/\au)] \sim [27.8, -0.6]$. It is fully surrounded by `young' stars, and as discussed in \S\,\ref{sec:biases}, most likely not shaped by the different exoplanet detection methods. Further, no upper limits are located on its right side, ensuring that no such limits might scatter into this gap. 

However, with the current dataset it is difficult to unambiguously interpret this feature as a result of X-ray driven photoevaporation. First of all, the small sample size reduces the voids' statistical significance. This is aggravated by the lack of knowledge of the expected void location, size and shape. In Appendix\,\ref{appendix}, we therefore discuss the significance of this void for two different cases: \textit{(i)} without any assumptions on its location/size (\S\,\ref{sec:no_apriori}) and \textit{(ii)} with an a priori knowledge of its location/size (e.g. assuming a numerical model could constrain these properties (\S\,\ref{sec:apriori}). This will allow us to forecast, how this analysis may benefit from additional observational data and more detailed numerical modelling.
For this purpose we have defined two reference regions, R1 and R2, in which our observational sample is assumed to be complete and not subject to any major observational biases. 

We find that the statistical significance of the observed void can only be proven in the current sample if its exact location and shape/size are known from theory. At this stage, our numerical model suffers from too many uncertainties which limits its predictive power. While an increased observational sample would certainly help to show the significance of any of the observed features, a more realistic numerical modelling is urgently needed to constrain the expected void location and size. 
One-dimensional approaches as those employed by AP12 and ER15 are plagued by a number of uncertainties and a systematic exploration of the result's sensitivity on the parameters used in the numerical simulations can help to better understand the underlying mechanisms. This will be the focus of a follow-up paper, which will solely aim at improving our theoretical understanding of how X-ray driven photoevaporation may affect the final semi-major axis distribution of giant planets.


\section{Summary and Conclusions}
\label{sec:summary}

We have searched for signatures of X-ray driven photoevaporation of planet-forming discs in the present-day semi-major axis distribution of giant planets. To that end we have constructed a catalogue, containing the X-ray properties of all known planet-hosting stars, which have been observed by \Chandra, \XMM\ and/or \Rosat. This catalogue contains basic stellar and planetary properties as well as X-ray fluxes and luminosities.

The main results from a statistical analysis of the data we collected combined with theoretical considerations can be summarised as follows:
\begin{itemize}

\item By correlating the X-ray luminosity of the host stars with the semi-major axis distribution of their giant planets, we found a prominent underdensity within the $\Lx$-vs-$a$ distribution, roughly centred on $(\Lx,a) \sim (10^{28}~\ergs,0.2~\au)$. To our knowledge, this void cannot be explained by any observational biases in the planet-detection process.

\item Due to the limited sample size of our observations, it is currently not possible to prove the significance of this void. Certainly, missions like TESS and eROSITA will help to resolve this issue by monitoring large parts of the sky, therefore increasing the sample of X-ray observations of giant planets hosts drastically.

\item{The possible void hinted in the observational data ($[\Lx,a] \sim [10^{28}~\ergs,0.2~\au]$) is at a different location from what is expected from simple theoretical considerations ($[\Lx,a] \sim [10^{29}$--$10^{30}~\ergs,1$--$2~\au]$). While the shift in $\Lx$ is readily explainable by considering the expected decay in X-ray luminosity of stars from Myr to Gyr ages, the shift in semi-major axis is more puzzling. In the case the statistical significance of this feature is confirmed, this would point to a clear knowledge of how high energy radiation from the stellar host affects the final architecture of giant exoplanets. }

\end{itemize}

\section*{Acknowledgements}

We thank Giovanni Rosotti and Jeff Jennings for helpful discussions and their support on using \texttt{SPOCK} for the numerical analysis. We also thank the anonymous referee for his/her extensive review, which led to the significant improvement of this paper.
We acknowledge the support of the DFG priority program SPP-1992 ``Exploring the Diversity of Extrasolar Planets'' (DFG PR 569/13-1, ER 685/7-1) \& the DFG Research Unit ``Transition Disks'' (FOR 2634/1, ER 685/8-1). 
B.E. acknowledges the support by the Munich Institute for Astro- and Particle Physics (MIAPP) of the DFG Cluster of Excellence `Origin and Structure of the Universe'. 
M.M.R. is supported by DOE grant DESC0011114.

The catalogue presented in this work is mainly based on stellar and planetary properties gathered from the \textit{Extrasolar Planets Encyclopaedia} \citep[\url{http://www.exoplanet.eu},][]{exoplanet.eu}.
This research has made use of data obtained from the $\it{Chandra}$ Data Archive and the $\it{Chandra}$ Source Catalogue, and software provided by the $\it{Chandra}$ X-ray Center (CXC) in the application packages \texttt{CIAO}, ChIPS, and Sherpa and of data obtained from the 3XMM \textit{XMM-Newton} serendipitous source catalogue compiled by the 10 institutes of the XMM-Newton Survey Science Centre selected by ESA.
This publication additionally makes use of data products from the Two Micron All Sky Survey, which is a joint project of the University of Massachusetts and the Infrared Processing and Analysis Center/California Institute of Technology, funded by the National Aeronautics and Space Administration and the National Science Foundation.
This work has made use of data from the European Space Agency (ESA) mission {\it Gaia} (\url{https://www.cosmos.esa.int/gaia}), processed by the {\it Gaia} Data Processing and Analysis Consortium (DPAC, \url{https://www.cosmos.esa.int/web/gaia/dpac/consortium}). Funding for the DPAC has been provided by national institutions, in particular the institutions participating in the {\it Gaia} Multilateral Agreement.\\

\textit{Software:} \texttt{AstroPy} \citep{astropy}, \texttt{CIAO} \citep{ciao}, \texttt{Matplotlib} \citep{matplotlib}, \texttt{NumPy} \citep{Numpy}, \texttt{SAOImage DS9} \citep{DS9}, \texttt{SciPy} \citep{scipy}, \texttt{SPOCK} \citep{ER15}, \texttt{TOPCAT} \citep{Topcat}


\bibliographystyle{yahapj}
\bibliography{literature} 


\appendix
\section{Statistical Analysis}
\label{appendix}

In \S\,\ref{sec:obs_vs_theory} we discussed that the void located at $[\log(\Lx/\ergs),\log (a/\au)] \sim [27.8, -0.6]$ in Figure~\ref{fig:results} cannot be readily explained by any observational biases. A void of similar shape and size is found in a simplified population synthesis model which we have performed using the same code and setup as \citet{ER15} \& \citet{Jennings+2018}, though it is located at lower X-ray luminosities and semi-major axes. While this shift between the observations ($\sim \Gyr$) and the simulation ($\sim \Myr$) can be somewhat qualitatively explained, it is currently not possible to predict the shifts' extent with high accuracy. This weakens the voids' significance within the observations, as either a larger observational sample or more detailed numerical models are required to obtain statistically significant results.
Therefore, we will investigate the void significance for two different cases in the following, namely \textit{(i)} without any assumptions on its location/size (\S\,\ref{sec:no_apriori}) and \textit{(ii)} with an a priori knowledge of its location/size by assuming our preliminary numerical model is constraining these properties (\S\,\ref{sec:apriori}). However, we restrict ourselves to 'young' stars only, as they are the only category for which a memory of the primordial X-ray luminosity is expected (cf. \S\,\ref{sec:xray_evolution}).

\subsection{No a-priori knowledge of the location/size of the void}
\label{sec:no_apriori}

Considering a randomly populated area of size $A_{\rm tot}$ and mean point density $n=N_{\mathrm{tot}}/A_{\rm tot}$, where $N_{\mathrm{tot}}$ is the total amount of data points in our reference region $A_{\rm tot}$, the probability to find a void of size $A_{\mathrm{void}}$ is described by the Poisson formula

\begin{equation}
P_0(nA_{\mathrm{void}}) = \exp{\left(-n A_{\mathrm{void}}\right)} \, .
\label{eq:poisson_distri}
\end{equation}
If one could accurately constrain the position of the underdensity from theory, this would yield the probability that the void is seen in a randomly distributed dataset using Eq.~\ref{eq:poisson_distri}. 
However, if the position of the underdensity cannot be constrained to high enough precision using simulations, we have to consider voids of arbitrary shape and position. 
The mean number of voids per unit area to be expected in a random distribution was derived by \citet{PolitzerPreskill1986} and \citet{Otto+1986} to be

\begin{equation}
D_0(n, A_{\mathrm{void}}) = \frac{(n A_{\mathrm{void}})^2}{A_{\mathrm{void}}} \exp{\left(-n A_{\mathrm{void}}\right)} \, .
\end{equation}
The prefactor to this equation takes into account that overlapping voids will be correlated instead of being independent from each other. We further note that the complexity of the tested region needs to be constrained to yield a reasonable analysis, since it will always be possible to find a large region of arbitrary, complex shape in a point cloud. For simplicity, we therefore assume a cubic test region with fixed horizontal orientation, noting that choosing a circular test region reduces the expected number of voids only slightly by $\approx 10\,\%$ \citep{Otto+1986}. We test against a random distribution in the $\log{(\Lx/\ergs)}$--$\log{(a/\au)}$ plane in which the void is most apparent. We consider two cases that correspond to different choices for the reference region ($A_{\mathrm{tot}}$) as illustrated in Figure~\ref{fig:results}. 
Since R1 neglects several systems with high X-ray luminosities and large semi-major axes, we increased the reference region in R2, in which only the clearly underpopulated region at high semi-major axes ($> 10~\au$) and low X-ray luminosities ($< 10^{28}~\ergs$) is omitted. 

\begin{figure}
\centering
\includegraphics[width=\columnwidth]{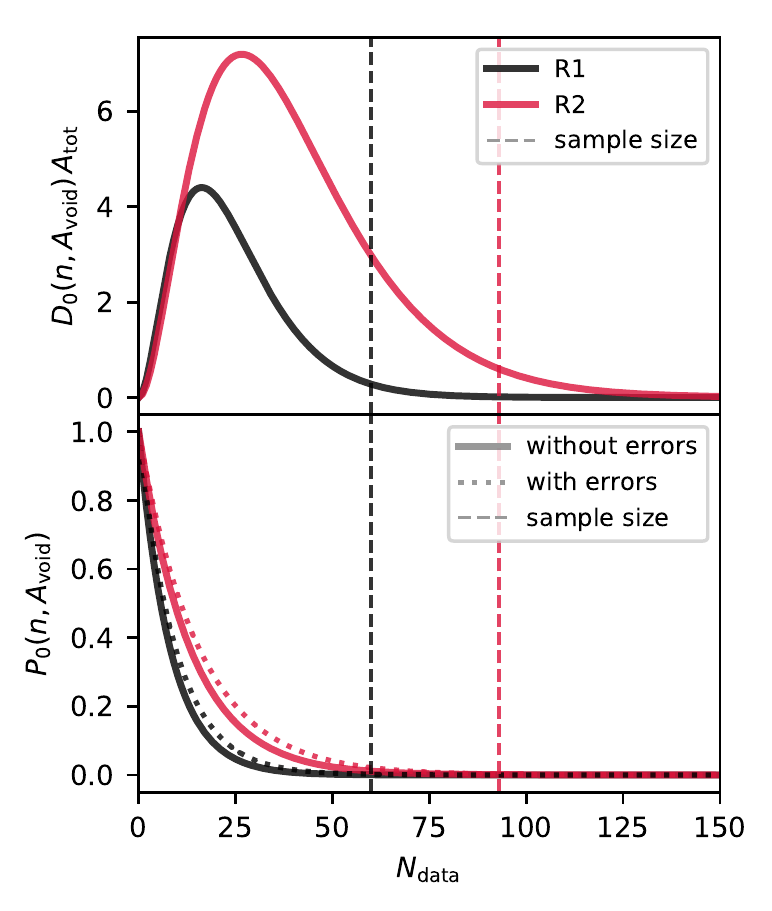}
\caption[]{\textit{Top}: Expected number of voids with fixed size and horizontal orientation, but variable position within the reference regions R1 and R2, respectively. The two solid curves show the two different scenarios for a range of possible sample sizes: only young stars (i.e. $< 2~\Gyr$) within R1 (black) and R2 (red), respectively. The vertical lines show the actual sample size of each test. \textit{Bottom}: The probability that the gap contains no data points without (solid lines) and with (dotted lines) including error measurements. The colours correspond to the same cases as in the upper panel.} 
 \label{fig:mc_result}
\end{figure}

The top panel of Figure~\ref{fig:mc_result} shows the expected number of voids for a random distribution $D_0(n, A_{\rm void}) \, A_{\rm tot}$, where $A_{\mathrm{tot}}$ describes the area of the reference regions R1 or R2, respectively. The vertical lines correspond to the true sample size of each selection.
Leaving the position of the void free and calculating the expected number of voids in a random distribution illustrates that for the current sample sizes, voids of the sizes considered in this study are to be expected. Thus, while the discussion in the previous section hints towards a physical origin of the observed void, with the current data its statistical significance cannot be shown. A significant increase in the sample size would be needed to allow a more detailed analysis of the presumed observational signature, which can help to refine our theoretical understanding of X-ray driven photoevaporation.

\subsection{A-priori knowledge of the location/size of the void}
\label{sec:apriori}

We now consider the case that the location and size of the void is known from theory. We show in the lower panel of Figure~\ref{fig:mc_result}
the probability of finding an empty, circular region in a random distribution of sample size $N_{\rm data}$ at the observed position for the two different scenarios. This test investigates whether the actual number of data points that are located within the reference regions, R1 and R2, is large enough to draw any conclusions on the significance of the presumed void.
The probability of finding the selected void with a priori knowledge of its location and size approaches zero for increasing $N_{\rm data}$, meaning that for larger sample sizes the probability of the gap being a result of random fluctuations approaches zero. The fact that the true sample sizes all lie in the flat region of each of these curves shows that the currently available amount of data on the X-ray properties of giant planet hosts is sufficiently large to interpret this void as significant, \textit{if} our numerical model can constrain its location and size. 

We further note the measurement errors of points near the corresponding void (assuming its location and size is well-constrained) imply that there is a small chance that some of them would be placed inside the underdensity. 
For simplicity, we therefore make the simple assumption that the measurement errors reduce the radius $R_{\mathrm{void}}$ of the void by $R_{\mathrm{void}}-R_{\mathrm{err}}$, where $R_{\mathrm{err}}$ is given by the largest error of those points that could scatter into the underdensity for the respective subsample. The resulting probabilities of finding this void of reduced size at the same location for each scenario as discussed before are shown as dotted lines in the lower panel of Figure~\ref{fig:mc_result}. Their deviation to the curves that do not take the errors into account is small and can be neglected for simplicity.

We finally note that the approach presented here considers uniformly distributed samples in the $\log{\Lx}$--$\log{a}$ plane as the random distribution. This sampling does therefore not correspond to a uniform sampling in linear $\Lx$--$a$ space, as the density of uniform samples drawn in logarithmic space, as viewed in linear space, is exponentially higher for lower values than for larger ones.

We conclude that it is currently not possible to prove the significance of the presumed void, as its exact location and shape/size is not known a priori and only roughly constrained by our preliminary theoretical investigation. An increased observational sample and/or a more sophisticated approach to the numerical modelling may in the future help show the significance of the presumed void, presenting a novel and direct way to explore the effects of disc dispersal on the final architecture of planetary systems.


\bsp	
\label{lastpage}
\end{document}